\documentclass{article}

\usepackage[english]{babel}
\usepackage[a4paper,top=2cm,bottom=2cm,left=3cm,right=3cm,marginparwidth=1.75cm]{geometry}
\usepackage{amssymb}
\usepackage{amsmath}
\usepackage[colorlinks=true,urlcolor=blue,anchorcolor=blue,citecolor=blue,filecolor=blue,linkcolor=blue,menucolor=blue,pagecolor=blue,linktocpage=true,pdfa=true]{hyperref}
\usepackage[utf8]{inputenc}
\usepackage[para]{footmisc} 
\usepackage{cite}
\usepackage{xspace}
\usepackage{url}
\usepackage{authblk}
\usepackage{graphicx}
\usepackage{xcolor}
\usepackage{orcidlink}

\author{Petr Balek\footnote{petr.balek@cern.ch}\ \orcidlink{0000-0002-0942-1966}}
\author{Tomasz Bo{\l}d\ \orcidlink{0000-0002-2432-411X}}
\author{Micha{\l} Naworyta\ \orcidlink{0009-0002-9017-4305}}

\affil{AGH University of Krakow, al.~Mickiewicza 30, 30-059 Krak\'ow, Poland}
\date{}

\title{The application of Hough transform for fast interaction vertex position estimation in heavy-ion collisions}

\newcommand{\MeV}{Me\kern -0.1em V\xspace}
\newcommand{\TeV}{Te\kern -0.1em V\xspace}
\newcommand{\sqn}{\sqrt{s_{_{\rm NN}}}}

\begin{document}
\maketitle
\begin{abstract}
Charged particle track reconstruction in silicon detectors of collider experiments in high-multiplicity events, such as heavy-ion collisions at LHC, is a difficult and resource-demanding process. The first phase of the procedure is the formation of seeds composed out of a few signals per track. A high number of actual particles results in a combinatorial explosion of the number of seeds. A priori knowledge of the collision vertex position would allow discarding non-viable track seeds early in the reconstruction procedure, reducing the overall computing requirements for the track reconstruction. The method proposed in this paper uses the Hough transform for estimating the position of the interaction vertex without the necessity of reconstructing the tracks first. It offers admissible resolution with linear scaling of numerical complexity with the multiplicity of tracks.

\vspace{0.4cm}

\end{abstract}


\section{Introduction}
\label{sec:intro}

Ultrarelativistic heavy-ion collisions, such as those occurring at the LHC, RHIC, or FAIR, generate a large number of charged particles. 
The precise reconstruction of the particles' trajectories is a prerequisite to virtually any physical measurement.
In ultra-central collisions, where the overlap of the colliding nuclei is maximal, the number of charged particles produced in a single collision can reach a couple of thousand per pseudorapidity unit~\cite{ALICE:2015juo, ATLAS:2011ag, CMS:2019gzk}. 

Reconstruction of the trajectories of these charged particles from measurements in a silicon detector is a difficult and time-consuming task. 
The required time is a non-linear function of a number of detected signals and can be as high as several hours per event if reconstruction of low transverse momenta (e.g.~100\,\MeV) of particles is required. 

One of the first stages of reconstruction involves the generation of the so-called seeds from the measurements left by the particles in the detector layers~\cite{ACTS}.
The seeds are typically doublets or triplets of measurements.
During seed generation, no assumption on vertex position along the $z$-axis is made; assumptions on the properties of the trajectories are very loose.
The number of seeds thus massively exceeds the actual number of particles in the given collision. This approach is typically taken in order to obtain the highest achievable efficiency, as missing the seed would inevitably lead to a missed track.

The luminous regions of the experiments are typically of micrometer size in direction transverse to the collided particle beams but span several centimetres along it. The actual collision vertex position is therefore changing from event to event along the beam direction (referred to further as $z$ direction).  Knowing the position of the vertex along $z$, even approximately, would allow us to put constraints on the origin of the charged particle tracks and thus consider only seeds originating from that estimated vertex. 
Such a constraint would significantly reduce the number of seeds to examine and consequently accelerate the entire reconstruction process. 
Previously suggested algorithms use combinations of detector hits (doublets or triplets) to generate vertex position estimates. 

In an experimental setup specific for large experiments like ALICE, ATLAS, or CMS, the magnetic field configuration is such that the induction vector $\vec{B}$ is parallel to the beam axis and relatively uniform. 
Then, a particle follows a helical trajectory, with the helix axis parallel to the beam axis.
If its transverse momentum is high enough, a line is a good approximation of the initial part of the trajectory projection in the $r$--$z$ subspace, where $r$ is the radial distance from the $z$-axis. 
These lines can either be extrapolated to the luminous region to find a maximum along the $z$-axis~\cite{Konstantinidis:2002pxa}, or the point of the closest approach to all these lines can be found. 
In either case, the found position would coincide with the collision vertex. 
The drawback of this approach is that the algorithm's complexity is polynomial as a function of the number of measurements. 

The approach suggested here employs the Hough transform (HT)~\cite{ht}. The Hough transform has been used for a long time to reconstruct charged particle tracks~\cite{HT_overview}, and is still used today in high-energy physics experiments, either for general track reconstruction~\cite{HT_NA61,HT_LHCb,HT_LHCb_2} or specifically for muon reconstruction~\cite{HT_ATLAS}.

The suggested approach has linear numerical complexity as a function of the number of measurements.
The studies of this method are performed using ACTS toolkit~\cite{ACTS} with the ODD detector model~\cite{ODD}, a simplified version of silicon tracking detectors to be used by the LHC experiments in High-Luminosity LHC phase~\cite{ATLAS:2017svb, ATLAS:2017azf, CMS:2017lum}. The Pb+Pb events are generated using \textsc{Pythia} Angantyr heavy-ion collisions model~\cite{Angantyr} at centre-of-mass energy per nucleon $\sqn = 5$\,\TeV.

This paper has 6 sections. 
The second section describes the method;
the third section discusses various parameters of the method and their impact on the resolution of the vertex position estimate and on the computational time demands;
the fourth section presents performance, using parameters tuned for the ODD;
the fifth section evaluates benefits of the additional seed filtering using the vertex position estimate;
and the sixth section offers a conclusion and outlook for future development.


\section{Method}
\label{sec:method}

In high-energy physics experiments, two types of silicon tracking detectors are typically used: pixel detectors and strip detectors. 
Each type has specific advantages and is utilised based on the requirements of the experiment.
The proposed algorithm demands precise information about the $z$-position of the measurements and thus is only applicable when pixel detectors are used. 
Also, in this particular implementation it can only be applied in the case of a uniform magnetic field aligned with the direction of the colliding beams. 
A complete absence of the field would also be admissible.

In case of a uniform magnetic field, the helical trajectories of the particles keep the angle $\theta$, angle between the trajectory and the $\vec{B}$ field, constant. 
Specifically, the angle $\theta$ is the same at the positions of all the measurements along the trajectory.
On the other hand, the angle $\theta$ depends on the position of the origin of the particle.

In line with typical HT implementations, the measurement coordinates are first transformed into a Hough image space. In the proposed method, the image space is spanned by two variables, $z_{vertex}$ which is the assumed vertex position and $\cot{\theta}$ using the formula:
\begin{equation}
    \cot(\theta) = (z_m-z_{vertex})/r_m,
\end{equation}
where $z_m$ and $r_m$ are cylindrical coordinates of a measurement..
Usage of $\cot(\theta)$ instead of $\theta$ reduces the computation to a single multiplication and addition per measurement and per $z_{vertex}$ assumption, making the computation very efficient.

\begin{figure*}[t]
    \centering
    \includegraphics[width=0.7\textwidth]{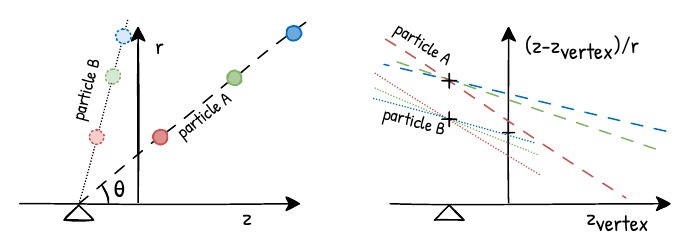}
    \caption{Sketch of the Hough transform. Left: Particles "A" (dashed) and "B" (dotted) have each three measurements: red, green, and blue. The angle $\theta$, between the particle trajectory and the $z$-axis, is generally different for each particle. The common vertex of origin is denoted by a triangle. Right: The same measurements are translated into lines in the Hough image space. They cross each other at the position $z_{vertex}$ and $\cot(\theta)$ of the associated particle. The values of $z_{vertex}$ are the same for both particles since they originate from the same vertex.}
    \label{fig:idea}
\end{figure*}

Figure~\ref{fig:idea} illustrates the construction of such an image space for three measurements originating from a single particle. 
The particle "A", represented by a dashed line, has 3 measurements. These three measurements are transformed into three dashed lines in the image space.
Analogically, the three measurements of the particle "B", represented by a dotted line, are transformed into three dotted lines in the image space.
The dashed (dotted) lines cross each other at the point determined by the vertex position and the $\cot(\theta)$ of the dashed (dotted) particle.
In collisions with several particles originating from the same vertex, the $\theta$ angles and consequently the $\cot(\theta)$ values would naturally differ from particle to particle, yet the $z_{vertex}$ position would be common, as is recognisable in the Figure~\ref{fig:idea} on the right.

Then, as is customary in HT, the image space is quantised (binned) into a 2D histogram, and its bins are filled each time they coincide with a line corresponding to a measurement.
After that, the bins of the histogram undergo a pedestal removal procedure. 
This process retains only the bins with a sufficient count of intersecting lines, enhancing the signal/background ratio of the image space.
After the pedestal removal, the histogram is integrated along the $\cot(\theta)$ dimension, transforming the 2D image space into a 1D histogram along the $z_{vertex}$ with a distinct peak.

\begin{figure*}[th]
    \centering
    \includegraphics[width=0.99\linewidth]{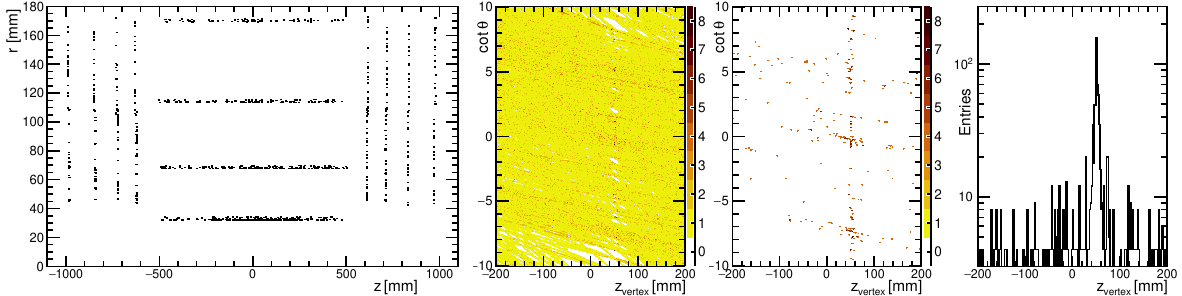}
    \caption{Example of the vertex estimation. From the left: Detector measurements projected into the $r$--$z$ space; Hough image space before the pedestal removal procedure; Hough image space after the pedestal removal procedure; and 1D projection of the image space along the $\cot(\theta)$ dimension with a clearly visible peak.}
    \label{fig:one-event-demo}
\end{figure*}

An example of this process is shown in Figure~\ref{fig:one-event-demo} for a simulated event with moderate particle multiplicity and a generated vertex at $ z_{vertex} = 50.9\,\text{mm}$. 
In this case, the initial image space might not show any evident pattern before pedestal removal. 
After the pedestal removal, a clear pattern emerges, which becomes even more pronounced in the 1D projection.

\section{Evaluation and algorithm parameters}
\label{sec:evaluation}

The effectiveness of the algorithm is measured by the resolution of the estimated vertex position. 
The algorithm shall perform equally effectively for events with large differences in occupancy, as is the case of heavy-ion collisions.

The performance of the algorithm depends on several parameters impacting how the image space is populated and treated: (a) the selection of the measurements considered for each $z_{vertex}$; (b) the granularity of the binning in the $\cot(\theta)$ dimension; (c) the granularity of the binning in the $z_{vertex}$ dimension.; (d) filling scheme for the image space histogram; and (e) the threshold on the number of entries in the HT image space in the pedestal removal procedure.

Optimal selection of these parameters is crucial for satisfactory algorithm performance.
Shall the parameters change from event to event, they should depend only on easily calculable quantities such as the total number of measurements or similar.

The effect of modifying the parameters on the 1D projection of the image space is shown in Figure~\ref{fig:eval}.

\subsection{Selection of the measurements}
\label{sec:etaselection}

In very high-multiplicity events, the high number of tracks results in an overpopulation of the image space. For that, an option to consider only part of the measurements was studied.
The set of measurements used for a given $z_{vertex}$ hypothesis is constrained to a wedge in $r-z$ space
with a certain opening angle, which is defined here using pseudorapidity $\eta = \text{arcsinh} [ \cot (\theta)]$.
An effect of constraining the $|\eta|$ range is shown in the upper left panel of Figure~\ref{fig:eval}. 
A wide range (large $|\eta|$ values) in high-occupancy events leads to many lines in the HT image space, and the consequent overpopulation precludes the peak finding.
On the contrary, a too-narrow range may result in the absence of a peak for events of small occupancy.

\subsection{Granularity of the image space along $\cot{\theta}$}

Another approach to tame the high occupancy of the HT image space histogram is to adjust its granularity along the $\cot(\theta)$-axis.
With finer granularity, it is less likely that several lines accidentally cross in the same bin. On the other hand, with too fine granularity the intrinsic uncertainties of the measurements positions
may result in lines caused by measurements of the same particle track that do not populate a common bin.

\begin{figure*}[tb]
    \centering
    \includegraphics[width=1\linewidth]{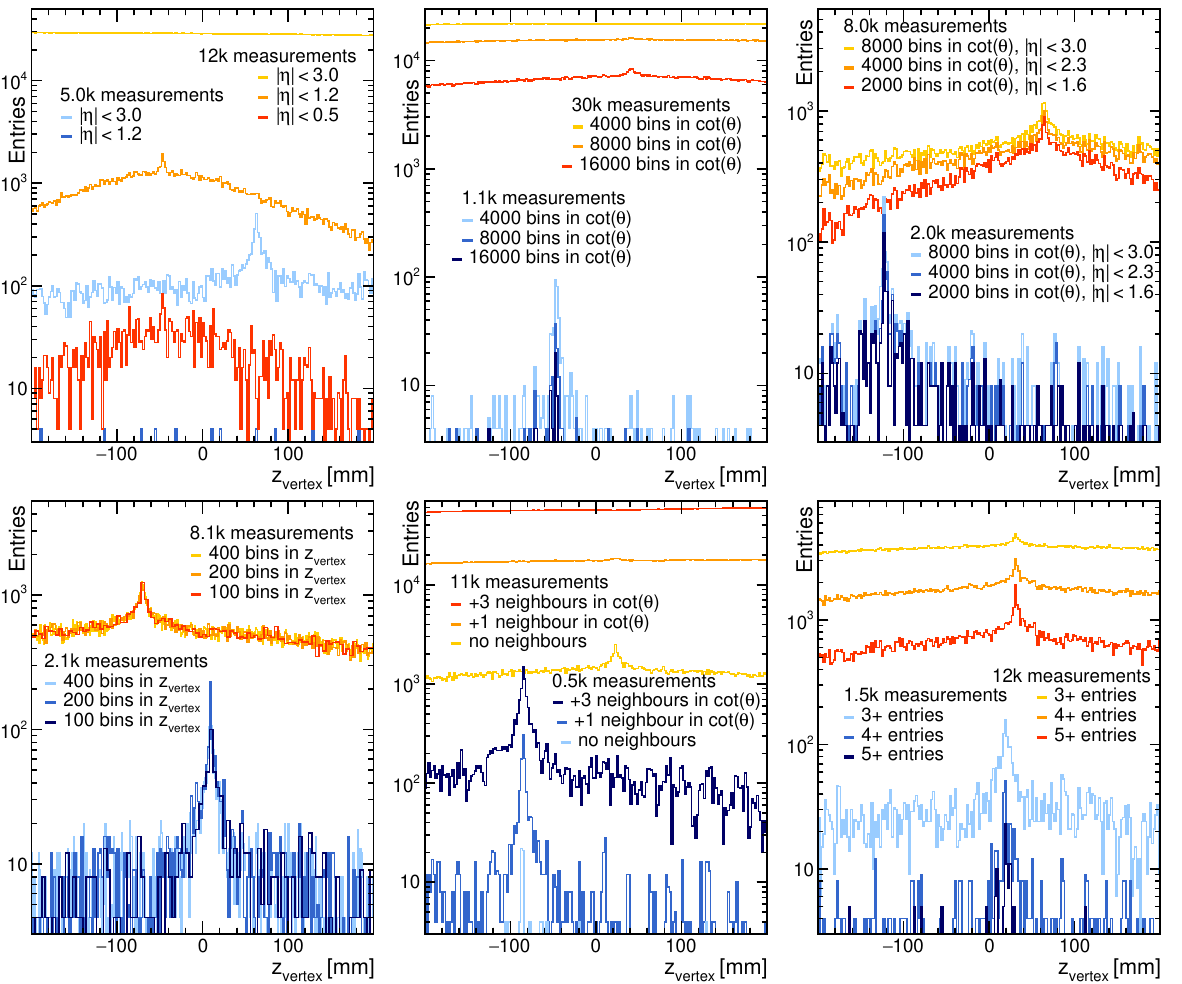}
    \caption{Exemplified effects of the different parameter setups for collisions with various multiplicities.
    Note different $y$-axis ranges on each panel. 
    Top row from the left: 
    Different considered $|\eta|$ range of the measurements; 
    different number of bins along $\cot(\theta)$-axis; 
    different number of bins along the $\cot(\theta)$-axis and different considered $|\eta|$ range such that the number of bins per unit of $\cot(\theta)$ remains the same (pseudorapidities $\eta = \pm3.0$, $\pm2.3$, and $\pm1.6$ are equal to $\cot(\theta) = \pm10.0$, $\pm4.93$, and $\pm2.38$, respectively, and thus utilising about 400 bins per unit of $\cot(\theta)$).
    Bottom row from the left:
    different number of bins along the $z_{vertex}$-axis;
    filling more bins in the HT image space along the $\cot(\theta)$-axis; and
    different minimum numbers of entries per bin in the pedestal removal procedure.
    Unless stated otherwise in the legend, the algorithm considers measurements within $|\eta|<3.0$, has 8000 bins along the $\cot(\theta)$-axis, has 200 bins along the $z_{vertex}$-axis, does not fill any neighbouring bins, and requires at least 4 entries per bin.}
    \label{fig:eval}
\end{figure*}

In this case, the pedestal removal procedure may eliminate a valid candidate.
As shown in the upper middle panel of Figure~\ref{fig:eval},
finer granularity may prevent peak formation in low-multiplicity events and coarse granularity results in overpopulation in high-multiplicity events.

The combined effect of adapting the $\cot(\theta)$ binning and $|\eta|$ range is illustrated in the top right panel of Figure~\ref{fig:eval}. 
As demonstrated, a fixed granularity per unit of $\cot(\theta)$ preserves a peak for diverse classes of events, hinting at a possible solution of the under- and overpopulation of the HT image space presented in the two previous panels.

\subsection{Granularity of the image space along $z_{vertex}$} 

The binning along the $z_{vertex}$-axis was also studied as a possible dynamic algorithm parameter.
It was found that it affects the performance only weakly, as shown in the bottom left panel of Figure~\ref{fig:eval}. The position and visibility of the peak are comparable for the number of bins changing by a factor of 2 or 4. Nevertheless, it has to be carefully considered due to its direct impact on computational time and memory consumption.

\subsection{Filling scheme for the image space} 

In experimental reality, the charged particles tracks even in uniform magnetic field do not form perfect trajectories due to scattering on detector material, sensors and support structures. 
Consequently, the bins in the HT image space must be sufficiently broad to accommodate this effect. This, on the other hand, would lead to overpopulation of the bins and deteriorate the performance in events of high occupancy. 
To recover the ``near-miss'' entries, the line corresponding to each measurement is widened, i.e.~the adjacent bins in $\cot(\theta)$ are also filled.
The bottom middle panel of Figure~\ref{fig:eval} shows the effect of such width expansion for events of different occupancy.
For very low-occupancy events, the wider lines help to identify the peak, while for ``bussier'' events, they obscure the desired result. 

\subsection{Pedestal removal} 

With a performant implementation of peak finding in mind, only a simple thresholding approach was studied.
The impact of the threshold value in the pedestal removal procedure is shown in the bottom right panel of Figure~\ref{fig:eval}. 
Requiring a high count per bin improves the significance of the peak in events with high occupancy; however, a too-high value results in suppression of the peak in events with lower occupancy.

\subsection{Memory \& computational time}

To ensure practical implementation, the design of the algorithm must consider the memory required to store the HT image space histogram and the computational time required to fill it.
The memory needed is proportional to the product of the number of bins in the $\cot(\theta)$- and $z_{vertex}$-axes. The time needed is proportional to the number of bins only in the $z_{vertex}$-axis.

To save both memory and computational time, the algorithm can be run multiple times, with increasing granularity (i.e.~more bins per mm) and decreasing range along the $z_{vertex}$-axis.

\subsection{Optimal parameters}
\label{sec:optimparams}

The following set of parameters was tuned to the ODD detector and is used to evaluate the algorithm performance discussed in Sections~\ref{sec:performance} and \ref{sec:effectiveness}.

If there are more than $10^4$ measurements in $|\eta|<3.0$, then the $|\eta|$ range is limited, so the utilised number of measurements remains close to $10^4$. 
The expected dependence of the number of measurements on $|\eta|$ is based on the average distribution of the measurements along $|\eta|$ over many collisions.
The actual number of measurements used for the estimation of the vertex in a particular event is not so important as long as it is not too high to cause overpopulation of the image space, as presented in Figure~\ref{fig:eval}.
Limiting the $|\eta|$ range is preferable to increasing the number of bins in $\cot(\theta)$ for two reasons. First, it limits the number of measurements used, reducing the processing time, and second, it limits the memory necessary for the HT image space histogram.

While adding the constraint on $|\eta|$ prevents over-population of the HT image space, the setup of other parameters wages against the resolution deterioration at lower-multiplicity events.

If there are at least $10^4$ measurements, then the $\cot(\theta)$-axis has 8000 bins.
For fewer measurements, the number of bins is reduced by a factor of $2^{4 - \log_{10} \left(N_{m}\right)}$, where $N_m$ is the number of measurements. This means that for every magnitude less, the number of bins decreases twice.
Moreover, if there are less than 1000 (200) measurements, then 1 (3) neighbours in $\cot(\theta)$ are filled as well. This helps with finding of the eventual peak as the tracks that underwent more scattering are effectively included.
Finally, the pedestal removal procedure requires at least 4 entries per bin, regardless of the number of measurements.

The computational time demanded by the algorithm is linearly proportional to the number of bins along the $z_{vertex}$-axis. To avoid running the algorithm over very fine-granularity wide-range axis, the algorithm is run three times: (1) with 800 bins in \mbox{$|z_{vertex}|<200$\,mm}; (2) with 180 bins in \mbox{$|z^{\prime}_{vertex}-z_{vertex}|<30$\,mm}; and (3) with 80 bins in \mbox{$|z^{\prime\prime}_{vertex}-z_{vertex}|<8$\,mm}.
Each iteration provides better resolution than the previous.


\section{Performance}
\label{sec:performance}

With the parameters optimised as described above, the performance was evaluated in 250k minimum-bias Pb+Pb events generated using \textsc{Pythia} Angantyr heavy-ion collisions model~\cite{Angantyr} with collision vertices chosen at random following the Gaussian distribution centred at \mbox{$(x,y,z) = (0,0,0)$\,mm} and width \mbox{$(\sigma_x, \sigma_y, \sigma_z) = (0.5,0.5,50)$\,mm}. Events feature wide range of multiplicities corresponding to varying collision centralities.

\begin{figure*}[th]
    \centering
    \includegraphics[width=0.99\linewidth]{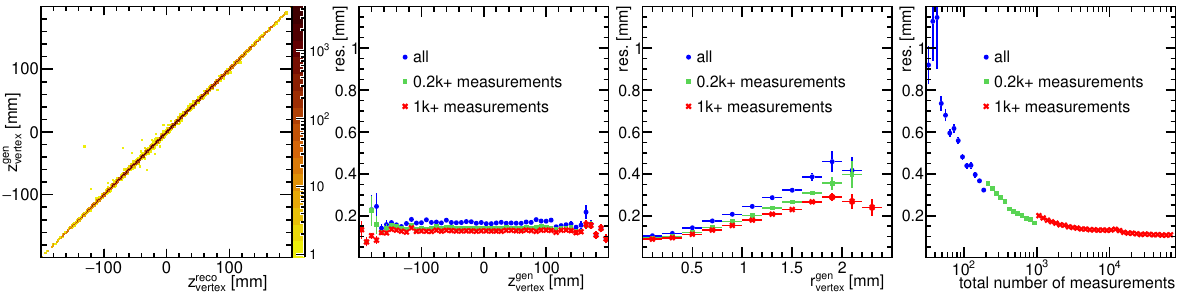}
    \caption{Performance of the algorithm. From the left: Correlation between reconstructed and generated positions of the vertices along the $z$-axis; resolution as a function of generated position along the $z$-axis; resolution as a function of generated position along the $r$-axis; and resolution as a function of measurements multiplicity in the event.}
    \label{fig:perf}
\end{figure*}

For every of 250k generated events, the vertex has been reconstructed successfully.
The leftmost panel of Figure~\ref{fig:perf} shows the correlation between the generated and reconstructed vertex positions.
An excellent correlation is found between these two quantities.
All event multiplicities are combined in this figure. 

The resolution, defined as the mean of a difference between the generated and reconstructed vertices in the $z$-direction, $\Delta z = |z_{vertex}^{reco} -z_{vertex}^{gen}|$, is shown for three event multiplicities as a function of the generated $z_{vertex}$-position, \mbox{$r$-position}, and the measurement multiplicity is shown in the three right panels of Figure~\ref{fig:perf}.

The resolution is flat along the $z$-axis. 
This is expected as the algorithm is agnostic about the position of the vertex along the $z$-axis.
Some deterioration at large $|z_{vertex}|$ may come from the detector design, though.
For the \mbox{$r$-position} of the vertex, the resolution worsens with increasing $r_{vertex}^{gen}$ because the algorithm assumes $(x_{vertex}^{gen},y_{vertex}^{gen})=(0,0)$ without any attempt to take such bias into account.
Finally, the resolution is better for high-multiplicity events, as the peak finding performs better on smoother 1D histograms. 

\section{Effectiveness}
\label{sec:effectiveness}

\begin{figure*}[ht]
    \centering
    \includegraphics[width=0.99\linewidth]{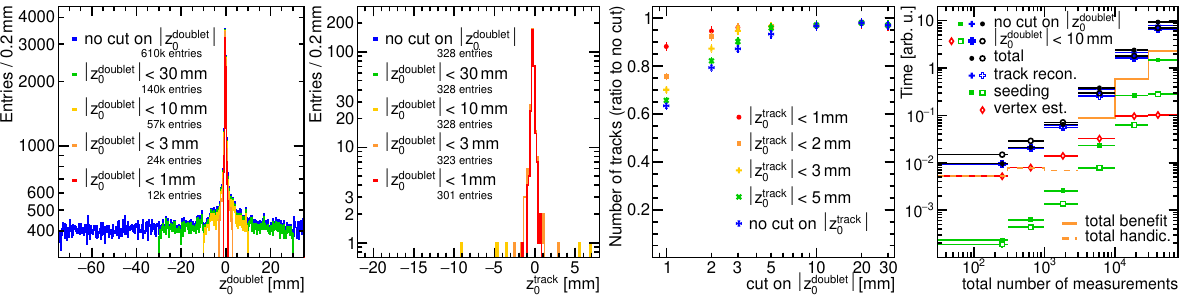}
    \caption{Track reconstruction with the additional seed filtering. From the left: Distribution of doublets as a function of longitudinal distance from the estimated vertex; distribution of tracks as a function of longitudinal distance from the estimated vertex; ratio of number of reconstructed tracks as a function of the $|z_{0}^{doublet}|$ cut to the number of reconstructed tracks without such cut; and computational time needed for the vertex estimation, seeding, and track reconstruction algorithms with 10\,mm cut on $|z_{0}^{doublet}|$ (open markers) and without such cut (closed markers). The overall time benefit (handicap) of the additional seed filtering is shown with the full (dashed) orange line. The horizontal positions of markers represent the mean multiplicities in each bin.}
    \label{fig:effect}
\end{figure*}

To evaluate the effectiveness of additional seed filtering using the estimated vertex position, 10k minimum-bias Pb+Pb events were generated with the same setup and vertex distribution as in the section~\ref{sec:performance}.

The two left panels of Figure~\ref{fig:effect} show an example of a single event with 10k measurements, $z^{gen}_{vertex}=20$\,mm, and $z^{reco}_{vertex}=20.2$\,mm. 
The actual restriction is imposed on doublets of measurements from which the seeds are created. The restriction on $z_{0}^{doublet}$, the distance of the doublet's closest approach to the estimated vertex along the $z$-axis, allows the rejection of doublets that point far away from the estimated vertex; this is shown in the leftmost panel. Thus, seeds pointing far away from the vertex are never created.

This rejection has almost no impact on the number of reconstructed tracks, as revealed in the second panel from the left, where several histograms overlap. The number of tracks is shown as a function of the track's closest approach to the estimated vertex along the $z$-axis.

The third panel from the left shows a ratio of the number of reconstructed tracks when using the cut on $|z_{0}^{doublet}|$ specified on the horizontal axis to the number of tracks reconstructed without any restriction on $|z_{0}^{doublet}|$. In both the numerator and denominator, only tracks within the $|z_{0}^{track}|$ range defined in the legend are considered.
A very tight constraint of a few mm in $|z_{0}^{doublet}|$ results in the loss of some reconstructed tracks. However, with a relaxed constraint of at least 10\,mm, the number of missed tracks is very small.
Furthermore, after applying the $|z_{0}^{track}|$ selection, as in a typical physics analysis, the loss of reconstructed tracks is negligible.

The right panel of Figure~\ref{fig:effect} shows the computational time needed for various stages of the track reconstruction as a function of the multiplicity of measurements. For events with fewer than approximately 3000 measurements, the extra time required by the vertex estimation algorithm is not compensated for by a faster reconstruction; the overall effect is negative.
It is expected that in the final implementation, events of low multiplicity will bypass the entire vertex estimation and additional seed filtering.
In events with more measurements, the extra time is compensated by faster seed and track reconstruction algorithms. The overall benefit is about 25\% of the total reconstruction time. The total number of seeds is also reduced by about 25\%.

Notably, the time needed by the vertex estimation algorithm remains approximately constant above $10^4$ measurements. This is due to the restrictions on the $|\eta|$ ranges, as mentioned in Sections~\ref{sec:etaselection} and \ref{sec:optimparams}.


\section{Summary}
\label{sec:summary}

The algorithm using the Hough transform for fast vertex estimation has been presented.
The numerical complexity of the algorithm is linear as a function of measurements for events with a wide range of charged-particle multiplicities and becomes sub-linear for high-multiplicity events.
The resolution is better for events with higher multiplicities, as long as the HT image space is not overpopulated.

For the ODD detector, the position estimate resolution is about 1\,mm for events with as little as 30 measurements; for events with more than 1000 measurements, the resolution is about $0.2$\,mm.
For even higher-multiplicity events, the resolution becomes even better.
Thus, usage of this algorithm in track reconstruction would not be severed when the reduction of computational time is most needed.
Similar performance is expected to be achieved for other detector designs.

A mild degradation of resolution is observed with increasing $r^{gen}_{vertex}$. It is expected that it can be recovered with an iterative scan with different assumptions of $(x_{vertex}^{gen},y_{vertex}^{gen})$.
Since it would be necessary to rerun the algorithm only for a narrow range in $z_{vertex}$, the impact on the computational time demands of the algorithm could be minimal.

Although events with multiple vertices were not investigated, it might be possible to detect more vertices, assuming that the binning along the $z$-axis is adequately fine and that the vertices are separated well enough. Moreover, the magnitude of the peak in the 1D projection might be an efficient estimate of overall event multiplicity usable in online filtering applications.
The size or shape of the peaks in the 1D projection could also serve for quick identification of a high-multiplicity vertex presence in the collisions. The algorithm applied to detector signals from hadronic jets in high pileup $pp$ collisions is expected to allow identification of the vertex of origin of each jet and thus allow for a fast identification of multi-jet vertices from hadronic decays of heavy particles.

An option is envisaged to employ a machine learning algorithm to identify features in the HT image space. Although this would undoubtedly increase the processing time, the overall effect might still be positive.


\section*{Acknowledgement}
\label{sec:ackn}

This work was supported by National Science Centre, Poland, under research project UMO-2023/51/B/ST2/00920.
The work of P.\,B.\ is supported by the programme ``Excellence initiative -- research university'' for AGH University of Krakow, grant no.~9041.


\bibliographystyle{atlasBibStyleWithTitle}
\bibliography{fast-vertex-reco}

\providecommand{\href}[2]{#2}\begingroup\raggedright\begin{thebibliography}{10}

\bibitem{ALICE:2015juo}
{ALICE Collaboration}, {\em {Centrality Dependence of the Charged-Particle Multiplicity Density at Midrapidity in Pb-Pb Collisions at $\sqrt{s_{_{\rm NN}}} = 5.02$\,TeV}}, \href{http://dx.doi.org/10.1103/PhysRevLett.116.222302}{Phys. Rev. Lett. {\bfseries 116} no.~22, (2016) 222302}, \href{http://arxiv.org/abs/1512.06104}{{\ttfamily arXiv:1512.06104 [nucl-ex]}}.

\bibitem{ATLAS:2011ag}
{ATLAS Collaboration}, {\em {Measurement of the centrality dependence of the charged particle pseudorapidity distribution in lead-lead collisions at $\sqrt{s_{_{\rm NN}}}=2.76$\,TeV with the ATLAS detector}}, \href{http://dx.doi.org/10.1016/j.physletb.2012.02.045}{Phys. Lett. B {\bfseries 710} (2012) 363--382}, \href{http://arxiv.org/abs/1108.6027}{{\ttfamily arXiv:1108.6027 [hep-ex]}}.

\bibitem{CMS:2019gzk}
{CMS Collaboration}, {\em {Pseudorapidity distributions of charged hadrons in xenon-xenon collisions at $\sqrt{s_{_{\rm NN}}}=5.44$\,TeV}}, \href{http://dx.doi.org/10.1016/j.physletb.2019.135049}{Phys. Lett. B {\bfseries 799} (2019) 135049}, \href{http://arxiv.org/abs/1902.03603}{{\ttfamily arXiv:1902.03603 [hep-ex]}}.

\bibitem{ACTS}
X.~Ai {et~al.}, {\em {A Common Tracking Software Project}}, \href{http://dx.doi.org/10.1007/s41781-021-00078-8}{Comput. Softw. Big Sci. {\bfseries 6} no.~1, (2022) 8}, \href{http://arxiv.org/abs/2106.13593}{{\ttfamily arXiv:2106.13593 [physics.ins-det]}}.

\bibitem{Konstantinidis:2002pxa}
N.~P. Konstantinidis and H.~Drevermann, {\em {Determination of the z position of primary interactions in ATLAS}}, CERN (2002).

\bibitem{ht}
P.~V.~C. Hough, {\em {Machine Analysis of Bubble Chamber Pictures}}, Conf. Proc. C {\bfseries 590914} (1959) 554--558.

\bibitem{HT_overview}
J.~Illingworth and J.~Kittler, {\em {A survey of the Hough transform}}, \href{http://dx.doi.org/https://doi.org/10.1016/S0734-189X(88)80033-1}{Computer Vision, Graphics, and Image Processing {\bfseries 44} no.~1, (1988) 87--116}.

\bibitem{HT_NA61}
{A. Merzlaya (for NA61/SHINE Collaboration)}, {\em {Track reconstruction in the inhomogeneous magnetic field for Vertex Detector of NA61/SHINE experiment at CERN SPS}}, \href{http://dx.doi.org/10.1088/1742-6596/798/1/012072}{Journal of Physics: Conference Series {\bfseries 798} no.~1, (2017) 012072}.

\bibitem{HT_LHCb}
{B. Storaci (for LHCb Collaboration)}, {\em {Optimization of the LHCb track reconstruction}}, \href{http://dx.doi.org/10.1088/1742-6596/664/7/072047}{Journal of Physics: Conference Series {\bfseries 664} no.~7, (2015) 072047}.

\bibitem{HT_LHCb_2}
L.~Calefice {et~al.}, {\em {Effect of the high-level trigger for detecting long-lived particles at LHCb}}, \href{http://dx.doi.org/10.3389/fdata.2022.1008737}{Frontiers in Big Data {\bfseries 5} (2022) 1008737}.

\bibitem{HT_ATLAS}
{ATLAS Collaboration}, {\em {Muon reconstruction performance of the ATLAS detector in proton{\textendash}proton collision data at $\sqrt{s}$ =13\,TeV}}, \href{http://dx.doi.org/10.1140/epjc/s10052-016-4120-y}{Eur. Phys. J. C {\bfseries 76} no.~5, (2016) 292}, \href{http://arxiv.org/abs/1603.05598}{{\ttfamily arXiv:1603.05598 [hep-ex]}}.

\bibitem{ODD}
P.~Gessinger-Befurt, A.~Salzburger, and J.~Niermann, {\em The Open Data Detector Tracking System}, \href{http://dx.doi.org/10.1088/1742-6596/2438/1/012110}{Journal of Physics: Conference Series {\bfseries 2438} no.~1, (2023) 012110}.

\bibitem{ATLAS:2017svb}
{ATLAS Collaboration}, {\em {Technical Design Report for the ATLAS Inner Tracker Pixel Detector}}, \href{http://dx.doi.org/10.17181/CERN.FOZZ.ZP3Q}{CERN (2017)}.

\bibitem{ATLAS:2017azf}
{ATLAS Collaboration}, {\em {Technical Design Report for the ATLAS Inner Tracker Strip Detector}}, CERN (2017).

\bibitem{CMS:2017lum}
{CMS Collaboration}, {\em {The Phase-2 Upgrade of the CMS Tracker}}, \href{http://dx.doi.org/10.17181/CERN.QZ28.FLHW}{CERN (2017)}.

\bibitem{Angantyr}
C.~Bierlich, G.~Gustafson, L.~L\"onnblad, and H.~Shah, {\em {The Angantyr model for heavy-ion collisions in \textsc{Pythia8}}}, \href{http://dx.doi.org/10.1007/JHEP10(2018)134}{JHEP {\bfseries 10} (2018) 134}, \href{http://arxiv.org/abs/1806.10820}{{\ttfamily arXiv:1806.10820 [hep-ph]}}.

\end{thebibliography}\endgroup

\end{document}